\documentclass{article}
\RequirePackage{fix-cm}
\usepackage{tabularx}
\usepackage{tikz}
\usepackage{amsmath}
\usepackage{listings}
\usepackage{graphicx}
\usepackage{xcolor}
\usepackage{fancyvrb}
\usepackage{authblk}
\usepackage[utf8]{inputenc}
\usepackage{hyperref}

\definecolor{codegreen}{rgb}{0,0.6,0}
\definecolor{codegray}{rgb}{0.5,0.5,0.5}
\definecolor{codepurple}{rgb}{0.58,0,0.82}
\definecolor{backcolour}{rgb}{0.95,0.95,0.92}

\def\BibTeX{{\rm B\kern-.05em{\sc i\kern-.025em b}\kern-.08em
    T\kern-.1667em\lower.7ex\hbox{E}\kern-.125emX}}

\lstdefinestyle{mystyle}{
    backgroundcolor=\color{backcolour},   
    commentstyle=\color{codegreen},
    keywordstyle=\color{magenta},
    numberstyle=\tiny\color{codegray},
    stringstyle=\color{codepurple},
    basicstyle=\ttfamily\footnotesize,
    breakatwhitespace=false,         
    breaklines=true,                 
    captionpos=b,                    
    keepspaces=true,                 
    numbers=left,                    
    numbersep=5pt,                  
    showspaces=false,                
    showstringspaces=false,
    showtabs=false,                  
    tabsize=2
}
\newcommand*\samethanks[1][\value{footnote}]{\footnotemark[#1]}

\title{On Iterative Evaluation and Enhancement of Code Quality Using GPT-4o}

\author[]{
    Rundong Liu\thanks{These authors contributed equally to this work}
    \
    André Frade\samethanks \ 
    \ 
    Amal Vaidya
    \
    Maxime Labonne
    \
    Marcus Kaiser
    \
    Bismayan Chakrabarti
    \
    Jonathan Budd\linebreak
    Sean Moran\thanks{Corresponding author: sean.j.moran@jpmchase.com}
}
\date{} 
\affil{JPMorgan Chase}

\begin{document}
\maketitle

\begin{abstract}
This paper introduces CodeQUEST, a novel framework leveraging Large Language Models (LLMs) to iteratively evaluate and enhance code quality across multiple dimensions, including readability, maintainability, efficiency, and security. The framework is divided into two main components: an Evaluator that assesses code quality across ten dimensions, providing both quantitative scores and qualitative summaries, and an Optimizer that iteratively improves the code based on the Evaluator's feedback. Our study demonstrates that CodeQUEST can effectively and robustly evaluate code quality, with its assessments aligning closely with established code quality metrics. Through a series of experiments using a curated dataset of Python and JavaScript examples, CodeQUEST demonstrated significant improvements in code quality, achieving a mean relative percentage improvement of 52.6\%. The framework's evaluations were validated against a set of proxy metrics comprising of Pylint Score, Radon Maintainability Index, and Bandit output logs, showing a meaningful correlation. This highlights the potential of LLMs in automating code quality evaluation and improvement processes, presenting a significant advancement toward enhancing software development practices. The code implementation of the framework is available at \href{https://github.com/jpmorganchase/CodeQuest}{https://github.com/jpmorganchase/CodeQuest}.
\end{abstract}

\section{Introduction}

The production of high-quality code remains a top priority for organizations developing software. High-quality code extends beyond syntactic or semantic correctness, encompassing attributes such as freedom from errors, readability, efficiency, portability, usability, testability, and maintainability~\cite{NDUKWE2023111524}. 

Evaluating code quality is a complex process due to the subjective nature of the concept, leading to the creation of various language-specific and  non-comprehensive tools that cover very specific aspects under the code quality concept umbrella~\cite{10.1145/1930464.1930480}. For example, the Python ecosystem provides security linters for identifying security vulnerabilities, style linters for identifying deviations from style guides, type checks, and error linters~\cite{Wilkes2020}. Moreover, subjective aspects like code design and readability are often guided by "best practices". It is also important to understand that evaluation of code quality, while crucial, is just one part of ensuring production-ready code. This is typically achieved through extensive code reviews, where multiple developers examine and modify the scripts until they meet the required standards. While important, this process is resource intensive, time-consuming, and it generally places additional burdens on senior developers that are ultimately responsible for vetting final code versions. Automating this process could significantly boost code development productivity.

Recent advancements in Large Language Model (LLM) technologies have shown promise in coding-related tasks, including code generation and evaluation~\cite{achiam2023gpt} \cite{spiess2024calibrationcorrectnesslanguagemodels} \cite{mahamud2023codequalityassessmentusing}. LLMs offer unique advantages for code evaluation due to their training on diverse programming languages and vast datasets, enabling them to first understand and then assess as well as improve code quality. Moreover, their ability to perform tasks based on prompts provides a flexible and highly configurable paradigm for easily defining complex tasks in natural language~\cite{kojima2023large}. Studies are increasingly focusing on how prompt engineering and complexity affect the LLMs' ability to generate accurate, maintainable code. For example, some research compares one-hot and iterative prompt strategies to understand how more detailed or structured prompts impact the functionality and readability of the generated code, showing that prompt complexity can significantly enhance the robustness of outputs~\cite{Angus_Yang}. Other studies assess GPT-4's performance against alternative models, highlighting notable improvements in debugging capabilities and responsiveness to follow-up prompts. Specifically, recent work investigates GPT's self-correction capacity when supplied with targeted feedback, such as error messages or static analysis results, highlighting the model's ability to refine code and address issues like smells or unnecessary complexity through iterative feedback~\cite{Przemyslaw} \cite{Liu_Yue}.

While LLMs' zero-shot capability have been widely exploited, recent studies~\cite{Shinn2023ReflexionLA}~\cite{anonymous2024enhancing}~\cite{zhang2024controllinglargelanguagemodelbased}~\cite{zhao-etal-2024-repair} have suggested that borrowing concepts from Reinforcement Learning could significantly improve the capability for LLM-based agents to perform reasoning and planning tasks. For instance, Zhang et al.~\cite{zhang2024controllinglargelanguagemodelbased} have drawn inspiration from the classical actor-critic method \cite{NIPS1999_6449f44a} to design a TipleCritic mechanism, which has significantly improved the success rate on multi-agent planning tasks.

Despite their potential, using LLMs for such complex tasks also presents challenges. LLMs are not inherently designed for quantitative assessments, which may affect the reliability of their scores~\cite{Hackl_2023}. Moreover, the probabilistic nature of LLMs can lead to inconsistencies. Indeed, small modifications to the prompt or a change of the LLM random seed can potentially lead to vastly different results~\cite{liu2023gpt} or ``hallucinations" which can be challenging to detect and control~\cite{maynez-etal-2020-faithfulness} \cite{huang2023survey}. Nonetheless, it is clear that LLMs hold immense potential to revolutionize code quality evaluation and improvement.

In this paper, we introduce CodeQUEST (Code Quality Understanding and Enhancement System Toolkit), an LLM-powered framework for assessing and improving code quality. CodeQUEST consists of an Evaluator and an Optimizer that work together through an iterative process to enhance the quality of code. The rest of the paper is organized as follows: Section~\ref{Methodology} introduces the CodeQUEST framework and its two components. Section~\ref{sec:experimental_setup} details the experimental setup. Section~\ref{results} presents results and insights on improving code quality using LLMs. Section~\ref{Threats} outlines threats to validity, and Section~\ref{Conclusion} concludes the discussion.

\section{The CodeQUEST Framework}\label{Methodology}

 The CodeQUEST framework consists of two components: an Evaluator and an Optimizer that leverage LLM to assess and improve code quality, respectively. In this work, we used the most recently released GPT-4o from the GPT-4 model series~\cite{achiam2023gpt} and code quality was defined across ten dimensions: \textit{Readability}, \textit{Maintainability}, \textit{Testability}, \textit{Efficiency}, \textit{Robustness}, \textit{Security}, \textit{Documentation}, \textit{Modularity}, \textit{Scalability}, and \textit{Portability}. 
 
\begin{figure}[t]
    \centering
    \resizebox{\linewidth}{!}{\usetikzlibrary{fit}

\definecolor{blue}{rgb}{0.9,0.96,1}
\definecolor{beige}{rgb}{1,0.96,0.90}
\definecolor{green}{rgb}{0.96,1,0.96}

\tikzstyle{label} = [align=center]
\tikzstyle{node_init} = [align=center, fill=blue, rectangle, minimum width=2cm, minimum height=1cm, text centered, draw=black]
\tikzstyle{node_improv} = [align=center, fill=beige, rectangle, minimum width=2cm, minimum height=1cm, text centered, draw=black]
\tikzstyle{node_final} = [align=center, fill=green, rectangle, minimum width=2cm, minimum height=1cm, text centered, draw=black]
\tikzstyle{init} = [rectangle, rounded corners, minimum width=3cm, minimum height=3.8cm, text centered, draw=black]
\tikzstyle{improvement} = [rectangle, rounded corners, minimum width=7cm, minimum height=3.8cm, text centered, draw=black]
\tikzstyle{arrow} = [->,>=stealth]

\begin{tikzpicture}[node distance=2cm]

\node (node0) [node_init] {Input\\code};
\node (node1) [node_init, xshift=2.4cm] {Evaluator\\assessment};
\node (node2) [node_improv, xshift=5cm] {Code quality\\improvement};
\node (node3) [node_improv, xshift=7.5cm] {Code\\validation};
\node (node4) [node_improv, xshift=10cm] {Evaluator\\assessment};
\node (node5) [node_final, xshift=12.5cm, yshift=0.7cm] {Final code};
\node (node6) [node_final, xshift=12.63cm, yshift=-0.7cm] {Final Evaluator\\ report};

\node (improvement) [improvement, fit={(node2) (node3) (node4)}] {};
\node (label_improvement) [label, yshift=1.63cm, xshift=7.35cm] { Optimizer};

\draw [arrow] (node0) -- (node1);
\draw [arrow] (node1) -- (node2);
\draw [arrow] (node2) -- (node3);
\draw [arrow] (node3) -- (node4);
\draw [arrow] (improvement) -- (node5);
\draw [arrow] (improvement) -- (node6);

\draw [arrow, bend left=45] (node4) to (node2);
\end{tikzpicture}}
    \caption{Schematic representation of CodeQUEST.}
    \label{fig:1-Agent-Flow}
\end{figure}
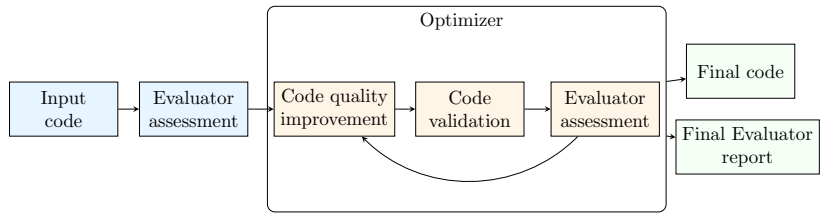

 The Evaluator first generates a code quality score\footnote{Under CodeQUEST framework, a code-level quality score is calculated by taking average across code quality score from each dimension} and a text-based evaluation for each dimension, which is used to produce an aggregated summary of dimension-wise evaluations. The Optimizer is then responsible for improving the code quality over a fixed number of iterations (or until a target quality score is reached) based on the Evaluator's feedback. 
 Each iteration involves three sequential stages: code quality improvement, code integrity validation, and code quality re-assessment. A detailed description of each stage is provided in this section below. A diagram of the end to end process can be found in Figure~\ref{fig:1-Agent-Flow}.

\subsection{Evaluator}\label{sec:code_quality_evaluation_framework}
As mentioned above, the Evaluator uses ten dimensions to assess code quality. Each dimension is addressed through a set of five questions or statements, carefully crafted to comprehensively cover its key aspects (c.f., Appendix~\ref{appendix_questions} for the full list). These questions and statements were designed to a) be general, thus applicable to most programming languages, and b) not overlap in dimension scope probing, hence avoiding over-representation of any particular aspects of the dimension. Furthermore, each question or statement is formulated such that it can only be answered with ``True", ``False" or ``Not Applicable", where ``True" reflects that high quality characteristics present in the code, in contrast to a ``False". We address each of the code quality dimensions in a separate query to the LLM, where the corresponding set of five dedicated questions or statements are provided to the LLM for its consideration.

Inspired by theoretical efforts~\cite{jiang2023latentspacetheoryemergent} \cite{tutunov2024largelanguagemodelsgenerate} on the impact of language ambiguity to LLM's performance, we carefully crafted prompts enabling LLMs to apply their internal knowledge in a focused and unambiguous manner. Limiting the possible output to quantifiable answers allows us to map the model output to a numerical scale.
The output includes the answers to each of the five dedicated questions or statements (which enable the quantitative assessment) and a high-level summary of the code in light of the dimension under consideration (which serves as a qualitative assessment).

Quantitative results are derived by assigning numerical values to each answer (+1 for ``True", -1 for ``False", and 0 for ``Not Applicable"). For each dimension, we sum up the five responses to obtain a score ranging from -5 to 5, where higher scores represent higher code quality along the dimension axis. Finally, dimension-specific scores are averaged to enable a ``code-level" score. We also ask GPT-4o to summarize qualitative assessment across all ten dimensions, obtaining a code-level summary.

The prompt template, based on zero-shot Chain-of-Thought (CoT)~\cite{cot}, is provided below. It combines 1) the code to be evaluated; 2) the set of five questions or statements for a given quality dimension; 3) the task to be performed, and 4) the desired output format:

\begin{Verbatim}
--- System ---
You are a helpful and harmless AI software engineer. 
You must provide an answer to the following request. 
Be brief and precise.

--- Human ---
### CODE: 
```
{code}
```

### STATEMENTS: 
{dimension_statements} 

### TASK:
Think step by step to assess the veracity of each STATEMENT 
in light of the CODE provided. 
Your answer to each statement must come from one of the following: 
* -1 if the statement is false, 
* 1 if the statement is true, 
* 0 if the statement is not applicable or there is not enough 
evidence in the CODE to address it. 

You must also provide a short summary about the quality of the
code from a {quality_dimension} perspective, justifying your 
answers across the various statements. 

### OUTPUT:
Return your answer in valid JSON as shown below:
```json
{{  "insight": <code quality summary:str>,
    "scores": [<score_to_statement1:int>, ...]
}}
```
\end{Verbatim}

Note that due to the inherent non-deterministic nature of GPT-4o (even when a seed is specified and the temperature is set to zero)~\cite{ouyang2023llm}, we allow users to opt-in self-consistency reasoning~\cite{Wang2022SelfConsistencyIC}\footnote{CodeQUEST framework allows user to apply self-consistency when evaluating each code quality dimension, a dimensional score is obtained by taking the average across retries and a dimensional evaluation is obtained by asking GPT-4o to summaries text evaluation across different retries} to further reduce the variance of the code quality score.

\subsection{Optimizer}\label{sec:code_quality_improvement_framework}

In this section, we describe the Optimizer, which is tasked with improving the quality of code based on the feedback generated by the Evaluator. While qualitative results set a clear direction for code improvement, quantitative feedback is leveraged to ensure that progress occurs in a unidirectional manner. The Optimizer operates in an iterative improvement cycle, where each iteration includes three steps: \emph{code quality improvement}, \emph{code validation}, and \emph{generated code evaluation}.

\subsubsection{Code Quality Improvement}

To generate an improved version of the code, the original code snippet and its qualitative assessment results are fed to GPT-4o. The task involves addressing all the areas of improvement identified in the assessment feedback through code modifications (the following prompt shares the system message with the Evaluator's prompt):

\begin{Verbatim}
--- Human ---
### Code:
```
{code}  
```
### Quality Dimensions Feedback: 
{quality_insight}  

### TASK: 
You are provided with a code script and detailed feedback 
for each quality dimension. 
For each quality dimension, you are provided with:
* A score from -5 to 5.The higher the score, 
    the better the quality.
* Dimension insights,highlighting potential areas 
    of improvement.
    
Think step by step to complete the following:
1) For each dimension, reflect on the score and insights.
2) Condense a list of improvement points, so that the code 
would be evaluated at a higher score for each dimension.
3) Improve the code script according to the improvement 
points, prioritizing dimensions with lower scores.
4) Return:
* the improvement points identified
* the improved version of the code script 
* explanations for each of the changes you've made
Note: 
* ALL improvement points MUST be addressed via meaningful 
changes to the code.
### OUTPUT: 
Your final output contains two parts:
Return your answer in a valid JSON as shown below:
```json
{{   
    "improvement_points": List[str],
    "explanation_report": List[str]
}}
```
Then quote your code in the following section:
```improved_code
{{improved_code_here}}
```
\end{Verbatim}

\subsubsection{Code Validation}

The code validation stage ensures that the LLM modified code can be compiled, as a minimal requirement for the code to be deemed valid and the process to proceed. Additionally, test cases built for the original code can be optionally executed against all code improvement versions to ensure the intended functionality and expected behavior. Such validation checks constitute a first step towards the meaningful evolution of the code. 

Note that the failure of either check leads to a rejection of the generated code as a candidate for the next iteration.  A failed attempt does however, still count as an iteration towards the total number of iterations.

Self-reflection and correction~\cite{Shinn2023ReflexionLA} based on the compiler or execution outputs were left as a interesting direction for future work.

\subsubsection{Evaluator Assessment}

A successful code validation stage is followed by a new Evaluator assessment. At this stage, if the overall quality score of the new code version drops relatively to the one of the previous version indicates that no overall improvement was achieved. Under these circumstances, a new iteration is triggered, using the inputs of the previous cycle as a new code improvement attempt. Again note that an unsuccessful improvement attempt still counts as an iteration. 

Otherwise, if the overall quality score of the new code increases relative to the last successful version, the iteration is deemed successful. In this case, the quality references are updated: the new overall quality score becomes the new numerical baseline for improvement and the qualitative feedback is used as input to the code quality improvement of the next cycle. 

The number of iterations and overall quality target score are both hyper-parameters of the Optimizer. The improvement cycle terminates when either, the maximum number of iterations or the target quality score is reached.

\section{Experimental Setup}\label{sec:experimental_setup}
\subsection{Dataset}\label{Methodology-Dataset}

We hand-curated a small but representative sample of 42 code examples across nine open-source libraries, including Scipy~\cite{2020SciPy-NMeth}, SecurityEval~\cite{siddiq2022seceval}, AWS-CDK-Examples~\cite{aws-cdk-examples} and Science-JS~\cite{scienc-js}. We focused on Python and JavaScript due to their popularity, and aimed for examples of varying code length and intended functionality. For future research, it would be interesting to expand the scope of the analysis to other programming languages, for example considering the QScored~\cite{sharma2021qscored} dataset focused on C\# and Java.

\begin{table}[th!]
    \centering
    \begin{tabular}{
        p{30mm}  
        p{18mm} 
        p{20mm} 
        p{20mm} 
        p{10mm}
    }
         \hline
         \textbf{Sources}    & \textbf{No. Examples}   & \textbf{Language}    &  \textbf{Test cases} & \textbf{No. Lines} \\ [0.5ex] 
         \hline
         
         auth0 \cite{auth0}      & 2          & JavaScript           & False                & 165 \\ 
         aws-cdk \cite{aws-cdk-examples}   & 2          & Python       & False               & 36 \\
         mbpp \cite{mbpp}     & 8          & Python       & True       & 19\\
         sciencejs \cite{scienc-js} & 6         & JavaScript            & False               & 112\\
         scipy \cite{2020SciPy-NMeth}     & 3         & Python      & False         & 117\\
         scikit-learn \cite{scikit-learn} & 3       & Python           & False    & 142 \\
         SnoopySecurity \cite{snoopysecurity} & 8      & Python, JavaScript       & False           & 60 \\ 
         fportantier \cite{fportantier} & 3         & Python       & False               & 39 \\ 
         SecurityEval \cite{siddiq2022seceval} & 7       & Python            & False           & 21 \\
         \hline
    \end{tabular}
    \caption{Detailed structural breakdown of the curated dataset.}
    \label{tab:dataset}
\end{table}

The majority of the code sourced from reputable open-source libraries should be of reasonably high quality. Thus, some of the code pieces were manually modified to enable scope for improvement. This included annotation removal and non-semantic changes to the code, whilst leaving the original logic intact. In Appendix~\ref{Appendix A}, we provide an example of reduced quality, where a utility function is re-defined multiple times, hence reducing the modularity of the code.

In this study we focused on 42 code examples (28 Python and 14 JavaScript), sourced from 9 different data sources. Table~\ref{tab:dataset} shows the complete list of data sources and associated metadata. Furthermore, test cases for 8 Python examples from MBPP~\cite{mbpp} were available, which we leveraged unchanged. For future research, it would be interesting to use LLMs to explicitly generate additional test cases~\cite{Shinn2023ReflexionLA}.

\subsection{Baseline solution} \label{Baseline}
To perform ablation study of the Evaluator's prompt, we considered a simpler CoT prompt to generate baseline results. This approach assumes that GPT-4o's internal knowledge is sufficient for an accurate and comprehensive evaluation of code quality. In our experiments, we compare it with CodeQUEST to validate the effectiveness of our solution. Again, the following prompt shares the system message with the Evaluator's prompt.

\begin{Verbatim}
---- Human ----
### CODE:
```
{code}
```

### TASK:
Think step by step to produce both a quantitative and 
qualitative assessment of the CODE provided. 

* Your qualitative assessment must be a short summary 
about the quality of the CODE.

* Your quantitative assessment must be an integer on 
a scale from -5 to 5, which respectively represent the
low and high-quality ends of the scale. 

Both types of evaluations must agree with each other.
    
### OUTPUT:
Return your answer in a valid JSON as shown below:
```json
{{
    "insight": <qualitative assessment:str>,
    "score": <quantitative assessment:int>
}}
```
\end{Verbatim}

\subsection{Validating Scores With Proxies}\label{sec:code_quality_score_validation}

Existing and well-established quality metrics are used in this study as a proxy to validate LLM-based code evaluations. This validation is limited to Python code due to the logistics of availability, implementation, and execution of such analytical tools. The following code quality evaluation capabilities were selected for this study as an attempt to cover as many code quality dimensions as possible:

\begin{enumerate}
    \item \textbf{Pylint}\footnote{We enabled all default Pylint extensions apart from mccabe, which computes the cyclomatic complexity to avoid partial overlap with Radon.}~\cite{Pylint} follows the style recommended by PEP8, while penalizing unused variables, lengthy code, and ill-named variables. Pylint returns a score from 0 to 10, which we use as a proxy metric for Readability and Documentation.
    \item \textbf{Radon Maintainability Index}~\cite{RadonMI} is used as a proxy for code Maintainability, similar to~\cite{Poldrack2023AIassistedCE}. The measure is a score from 0 to 100 derived from other metrics, including Source Lines of Code (SLoC), Cyclomatic Complexity, and Halstead Volume. We scale the score to range from 0 to 10.
    \item \textbf{Bandit}~\cite{Bandit} scans Python code for security issues for Python using abstract syntax trees. Bandit does not provide a score, so we applied a simple heuristic that transforms the execution standard output into a score from 0 to 10 to define our proxy for Security.
\end{enumerate}

For simplicity, we define our overall proxy code quality score as the average of these three metrics, resulting in a final score ranging from 0 to 10.

\section{Results}\label{results} 
\subsection{Can GPT-4o evaluate code quality?}\label{sec:compare-report}

To evaluate the CodeQUEST's Evaluator, we compare its evaluation results to those produced by the baseline setup. In both cases, the output consists of an overall code quality score (quantitative result) as well as a short summary report (qualitative feedback). Each code example was repeatedly evaluated five times by both setups\footnote{In this experiment, we disabled self-consistency for the CodeQUEST evaluator to ensure a fair comparison}, allowing us to gather statistics on the consistency of the outputs produced.

\begin{table}[!ht]
    \centering
    \begin{tabular}{|p{0.16\textwidth} | p{0.7 \textwidth} | p{0.05\textwidth}|}
        \hline
        Example & mbpp/601.py & Score\\
        \hline
        Baseline & The code is functional and correctly implements the logic to find the maximum chain length of pairs. However, it lacks comments, proper variable naming, and error handling. The use of 'max' as a variable name is not recommended as it shadows the built-in function 'max'. The code could be improved for readability and maintainability. & 2.0 \\
        \hline
        CodeQUEST & The provided code is a straightforward implementation of finding the maximum chain length of pairs using a dynamic programming approach. While the logic is correct, the code has several areas for improvement. It lacks readability and consistent style, with variable names that could be more descriptive. The use of 'max' as a variable name is misleading. The code does not include comments to explain the logic, especially the nested loops and the purpose of the 'mcl' array. It is moderately maintainable but could benefit from refactoring to improve efficiency and reduce repetition. The code is easy to test due to its simplicity and lack of external dependencies, but it does not facilitate mocking of dependencies. The nested loops result in $O(n^2)$ time complexity, making it inefficient for large datasets. The code lacks input validation, error handling, and concurrency considerations. It is not designed with scalability or resource efficiency in mind. Additionally, the code lacks documentation and modularity, making it difficult to test and modify parts independently. Despite these issues, the code is portable as it avoids platform-specific features and uses standard libraries. & -1.3\\
        \hline
    \end{tabular}
    \caption{Comparison of the qualitative and quantitative results with the baseline solution and CodeQUEST on mbpp/601.py.}
    \label{tab:mbpp_604_comparison}
\end{table}

\begin{figure}
    \centering
    \includegraphics[width=0.9\linewidth]{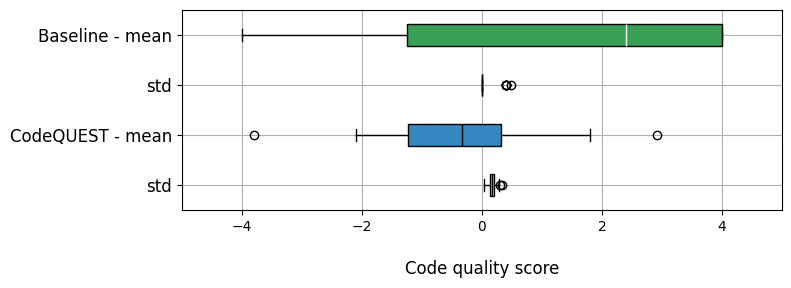}
    \caption{Comparison of the Baseline and CodeQUEST code quality scores.
    Each code example was evaluated five times. The mean and standard deviation were calculated across these five samples for the baseline (top) and CodeQUEST (bottom).}
    \label{fig:compare-box}
\end{figure}

\begin{itemize}
    \item[-] \textbf{Qualitative evaluation}: As illustrated in Table~\ref{tab:mbpp_604_comparison} and Appendix~\ref{appendix_code_quality_examples}, qualitative results produced by CodeQUEST were found to be more comprehensive and detailed compared to those of the baseline solution. For example, CodeQUEST Evaluator performed additional time complexity analysis against \textit{mbpp/601.py}, illustrating areas of improvement for Scalability.
    Nonetheless, both sets of results raise accurate points about the code, providing us with confidence that GPT-4o seems to have advanced intrinsic knowledge of code quality. We also found that the assessment produced equally plausible results for Python and JavaScript code, which showed that the LLM is able to generate insights across different programming languages.
    \item[-] \noindent\textbf{Quantitative evaluation}: As shown in Figure~\ref{fig:compare-box}, our two strategies exhibit different mean quality scores with some variability. The baseline appears to systematically overestimate code quality compared to CodeQUEST assessments. Since the deficiencies pointed out by the Evaluator are the primary signal for driving code improvement by the Optimizer, such overestimation of quality (both qualitative and quantitative) is undesirable, as it may hinder scope for improvement. For both setups, the variability may be explained by a) the ambiguity of prompt questions or statements, b) inherent LLM stochasticity, and c) the scale of possible outcomes not being granular enough.
\end{itemize}

Therefore, in summary, our results suggest that we can leverage GPT-4o for both qualitative and quantitative assessment of code quality. Furthermore, the accuracy and level of detail of LLM-based code evaluations are highly prompt-dependent. We also find that CodeQUEST provides a more comprehensive, standard, and reliable evaluation due to the additional guidance enabled by dimension-specific questions.

\subsection{Can evaluations be used to drive code quality improvement?}

In our study, we applied the CodeQUEST framework to each code example for a maximum of five improvement cycles. All versions of the Python code were checked for their ability to compile. Furthermore, all versions of the mbpp code were validated against the corresponding test cases provided. All intermediary and final results were recorded for downstream analysis.

\begin{figure}[!ht]
    \centering
    \includegraphics[width=1.0\linewidth]{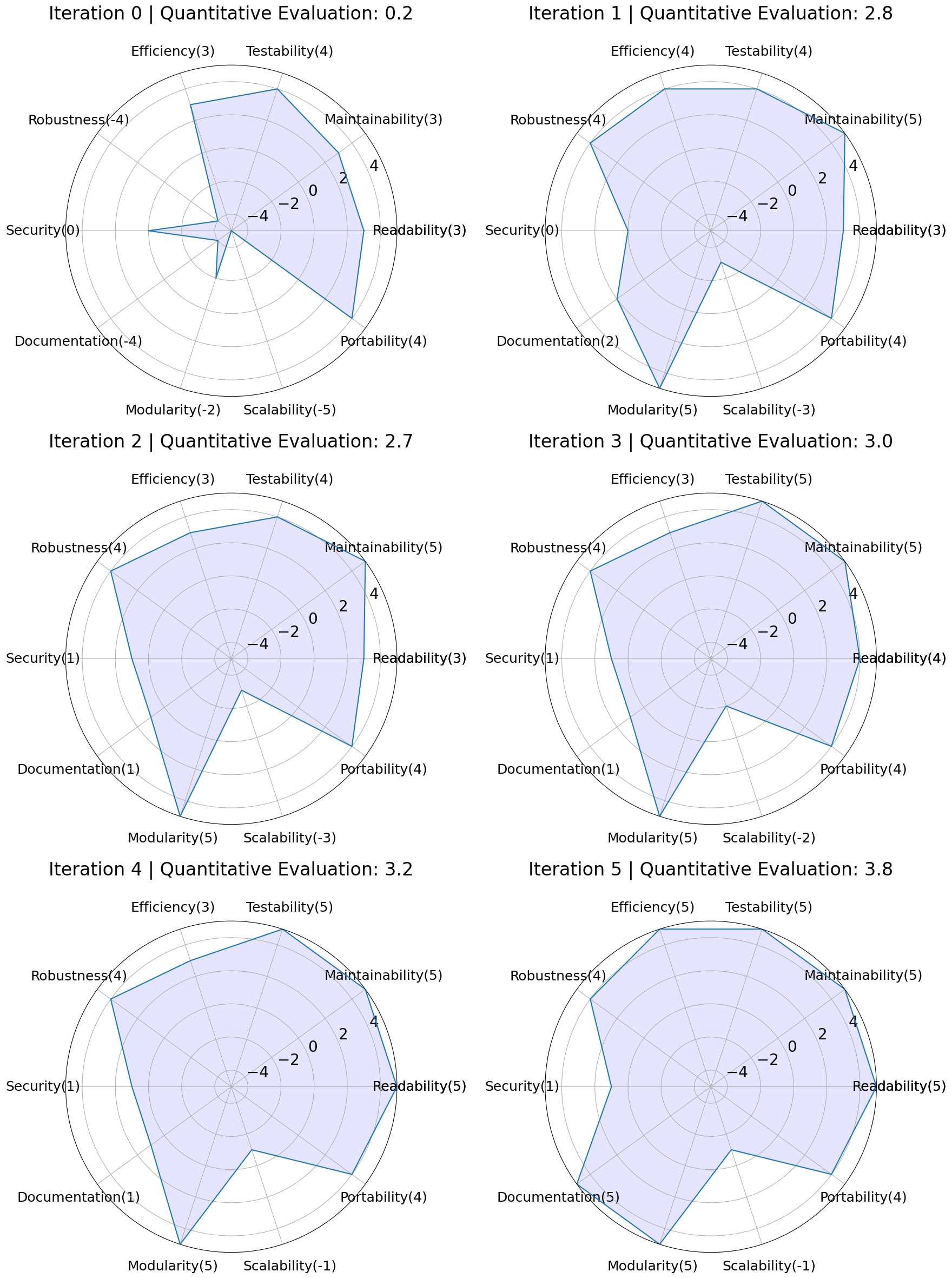}
    \caption{Quality score evolution of a Python code example (mbpp/927.py) when subjected to the Optimizer. The quality score is provided in its overall and dimension-specific format. Corresponding quality summaries and code snippets can be found in the Appendix~\ref{CodeQUEST - code evolution example}}
    \label{fig:examples_quality_improvement}
\end{figure}

The example shown in Figure~\ref{fig:examples_quality_improvement} represents the general behavior observed across all code examples. Out of the original 42 code instances, 41  were improved to some extent. We report an average absolute improvement of $2.9 \pm 1.3$ in the code quality scale units\footnote{As a by-product of this step, the original dataset was expanded with 208 additional code examples, excluding 2 invalid attempts to improve Python code} with those of initial lower quality exhibiting the largest improvements. 

To further analyze the improvements, we defined the {\it Relative Percentage Improvement(RPI)} as
\begin{equation}\label{eqn:mrpi}
    \operatorname{RPI} := 100\frac{(s^n_f - s^n_i)}{(s_{\max}-s^n_i)}
\end{equation}

\begin{figure}[!ht]
    \centering
    \includegraphics[width=0.9\linewidth]{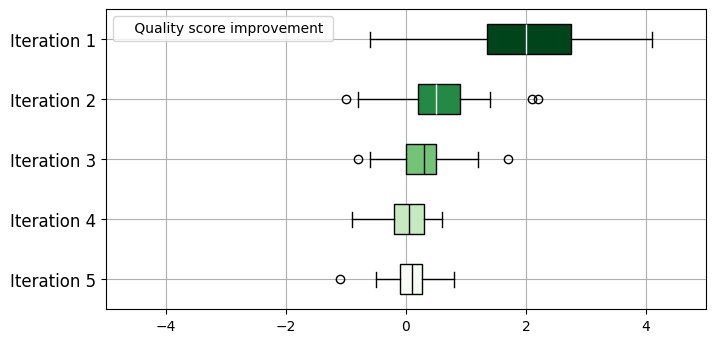}
    \caption{Distribution of code quality score improvements per iteration. (We disabled self-consistency in the experiment)}
    \label{fig:iterative_improvement_absolute}
\end{figure}

where $s_{\max}=5$ is the maximum possible score, and $s^n_i$ and $s^n_f$ are the initial and final scores for the $n^{th}$ code example, respectively. $N=42$ is the total number of examples in the dataset. Over the entire dataset, our framework produces a mean RPI of $52.6\% \pm 17.9\%$ and a median RPI of $57.1\%$.
We also verified that while our setup, by design, ensures a monotonic improvement of overall code quality, the monotonic improvement of individual dimensions' scores does not always occur, i.e.,~an overall improvement for the quality score produced by an iteration might be accompanied by a decrease in some of the dimension-specific scores.

Figure~\ref{fig:iterative_improvement_absolute} displays the distribution of improvements achieved for each iteration in absolute code quality units. We can observe that most of the improvement occurs in the first iteration cycle, with an average of 2 code quality units increase. The average improvement for iterations two and three were found to be 0.52 and 0.27. From the fourth iteration, the improvement becomes negligible (below 0.1 code quality units).

These results suggest that the magnitude of code quality improvement reduces at each iteration, which becomes particularly useful to limit the GPT-4o API costs incurred by the framework if CodeQUEST is to be considered and adopted at scale. 

Therefore this section demonstrates that our Optimizer framework can utilize the qualitative insights generated by the Evaluator to achieve significant code quality improvement. Indeed, we notice the consistent alignment between quantitative and qualitative results, such that an increase in the quality score reflects a decrease in code quality deficiencies reported by the quality summary. Therefore we believe that this setup should be general enough to treat this as a general-purpose recipe for configurable code quality improvement.

\subsection{Are these evaluations meaningful?}\label{sec:results:eval_meaningful}

To confirm the validity of LLM-generated quality scores, we investigate their level of agreement with the quality proxy scores described in Section~\ref{sec:code_quality_score_validation}.
Given each original code example and its corresponding refactored versions generated by CodeQUEST, here denoted as $\{x_0, \dots, x_{n_{max}}\}$, we can consider the incremental change of the score per iteration, defined as 
\begin{equation}\label{equation:2:delta scores}
    \Delta_{x_n}{\text{CodeQUEST}} = \text{CodeQUEST}(x_n) - \text{CodeQUEST}(x_{n-1}),
\end{equation}
for iterations $n=1, \dots, n_{max}$. In our case, $n_{max}=5$. Similarly, we can define $\Delta_{x_n}{\text{Proxy}}$ and $\Delta_{x_n}{\text{Baseline}}$ for the proxy and baseline scores, respectively.
\begin{table}[!ht]
    \centering
    \begin{tabular}{l c c c c c } 
         \hline
         \textbf{Pairs}  & $r_s$ & $r_p$ & $\mathrm p_s$ & $\mathrm p_p$\\ [0.5ex] 
         
         \hline 
         $(\Delta \text{Proxy}$, $\Delta \text{Baseline})$ & 0.21 & 0.27 & 0.02 & 0.00
         \\
         $(\Delta \text{Proxy}$, $\Delta \text{CodeQUEST})$ & \textbf{0.23} &\textbf{0.53} & \textbf{0.01} & \textbf{0.00} 
         \\
         \hline 
    \end{tabular}
    \caption{Spearman $r_s$ and Pearson $r_p$ correlation with corresponding p-values $\mathrm p_s$ and $\mathrm p_p.$}
    \label{proxy_correlation_table}
\end{table}
\begin{figure}[!ht]
    \centering
    \includegraphics[width=\linewidth]{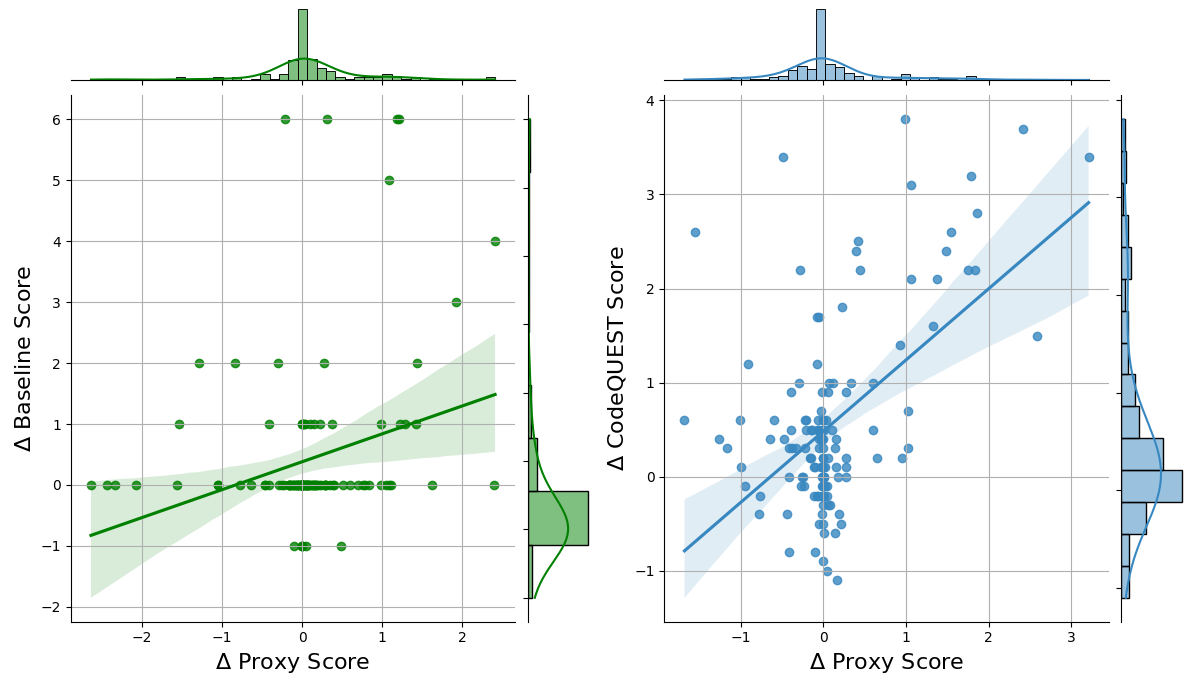}
    \caption{Left: Scatter plots and linear regression line for $\{\Delta_{x_n}{\text{Proxy}}$, $\Delta_{x_n}{\text{CodeQUEST}} \}_{x_n}$ (left) and $\{\Delta_{x_n}{\text{Proxy}}$, $\Delta_{x_n}{\text{Baseline}}\}_{x_n}$ (right). The shaded area represents the uncertainty of the linear regression fit.}
    \label{fig:delta_correlation}
\end{figure}
The iteration cycle applied to the 28 Python code examples described in Table~\ref{tab:dataset} resulted in 138 improved code examples\footnote{We set the maximum number of evolution cycle to be 5, we excluded 2 improvement attempts which were syntactically incorrect or failing the test case.} for which we calculated the incremental change for each of the three scores. To measure how much the LLM-based scores are aligned with proxy results, we compute the Pearson $r_p$, and Spearman rank $r_s$ correlation between $\Delta{\text{Proxy}}$ and  $\Delta{\text{CodeQUEST}}$ as well as  $\Delta{\text{Proxy}}$ and $\Delta{\text{Baseline}}$ and report the results in Table~\ref{proxy_correlation_table} and Figure~\ref{fig:delta_correlation}.

Results suggest that proxy scores are considerably better correlated with the CodeQUEST scores compared to those of the baseline. This suggests that, given a base code example, CodeQUEST is able to iteratively assign scores that better reflect the underlying variation in the code quality.
We can observe from Figure~\ref{fig:delta_correlation} that the relationship is relatively noisy. This is somewhat expected since the proxies do not capture all the dimensions considered by CodeQUEST and can be focused on certain aspects. For example, when we inspected individual results, we found that CodeQUEST is able to detect security vulnerabilities, which were ignored by Bandit. We provide two such examples from SecurityEval in Appendix~\ref{appendix_code_security_vulnerability_examples}, along with their improved version.

\section{Threats to validity}\label{Threats}

In this section, we would like to acknowledge certain considerations regarding the validity and robustness of our framework. Firstly, code quality is an intrinsically subjective concept and, while our evaluations work well with the selected questions, it might not be as effective if modified significantly. However, given the generalization capabilities of GPT-4o and its advanced knowledge of both code and English language concepts, we believe this is unlikely as long as the questions are reasonable and posed in the manner we detail in Section~\ref{sec:code_quality_evaluation_framework}.

Moreover, the stochasticity of GPT-4o output also introduces some inherent variability in the results. While we see good stability in our setup and take measures to limit it by setting the temperature to 0, that might not hold for all code and or prompts. We also recommend using the self-consistency framework~\cite{Wang2022SelfConsistencyIC} to address this, if required.

Hallucinations are also a well-known shortcoming of LLMs and this is an active field of research, thus it is possible that the model hallucinates during the improvement step of our framework. A detailed study with a larger dataset and human validation would be required to ascertain the extent to which hallucination might be a concern. We also recommend robust validation with test cases to guard against this. Our framework might also struggle to improve code that is very different from the training dataset of GPT-4o (e.g., highly custom proprietary code bases).

Finally, further study would be required to fully validate the framework across a broader range of coding languages. Here we note that the primary limitation comes from the LLM's (GPT-4o's) knowledge of the programming language, so the framework should be performant across most major programming languages.     

\section{Conclusion}\label{Conclusion}
In this paper, we introduced CodeQUEST -- Code Quality Understanding and Enhancement System Toolkit -- a  GPT-4o powered framework capable of evaluating and improving code quality for code across different domains. We show that our framework is able to produce comprehensive qualitative and quantitative code quality evaluation reports, and subsequently use these reports to iteratively improve the quality of the code being analyzed. Our framework is easily understandable, adjustable, and shows impressive results for Python as well as JavaScript, showing its applicability across languages.
We also demonstrate that our framework produces code evaluation scores showing good agreements with rule-based code-quality evaluation libraries.

\bibliographystyle{plain}
\bibliography{main}


\section{Conflict of interest}
The authors declare that they have no conflict of interest.
\section{Disclaimer}
This paper was prepared for informational purposes by the Applied Innovation of AI team of JPMorgan Chase \& Co. This paper is not a product of the Research Department of JPMorgan Chase \& Co. or its affiliates. Neither JPMorgan Chase \& Co. nor any of its affiliates makes any explicit or implied representation or warranty and none of them accept any liability in connection with this paper, including, without limitation, with respect to the completeness, accuracy, or reliability of the information contained herein and the potential legal, compliance, tax, or accounting effects thereof. This document is not intended as investment research or investment advice, or as a recommendation, offer, or solicitation for the purchase or sale of any security, financial instrument, financial product or service, or to be used in any way for evaluating the merits of participating in any transaction. The described work is a prototype and is not a production deployed system.

\appendix

\section{Manual code quality defect}\label{Appendix A}

Example of code that has been ``degraded in quality". We first show the original code piece, followed by the modified code, where the {\bf\_update\_x} function is re-defined multiple times.

Original code example from Scipy:
\begin{lstlisting}[language=Python]
class LinearVectorFunction:
    """Linear vector function and its derivatives.
    
    Defines a linear function F = A x, where x is N-D vector and
    A is m-by-n matrix. The Jacobian is constant and equals to A. The Hessian
    is identically zero and it is returned as a csr matrix.
    """
    def __init__(self, A, x0, sparse_jacobian):
        if sparse_jacobian or sparse_jacobian is None and sps.issparse(A):
            self.J = sps.csr_matrix(A)
            self.sparse_jacobian = True
        elif sps.issparse(A):
            self.J = A.toarray()
            self.sparse_jacobian = False
        else:
            # np.asarray makes sure A is ndarray and not matrix
            self.J = np.atleast_2d(np.asarray(A))
            self.sparse_jacobian = False
    
        self.m, self.n = self.J.shape
    
        self.xp = xp = array_namespace(x0)
        _x = atleast_nd(x0, ndim=1, xp=xp)
        _dtype = xp.float64
        if xp.isdtype(_x.dtype, "real floating"):
            _dtype = _x.dtype
    
        # promotes to floating
        self.x = xp.astype(_x, _dtype)
        self.x_dtype = _dtype
    
        self.f = self.J.dot(self.x)
        self.f_updated = True
    
        self.v = np.zeros(self.m, dtype=float)
        self.H = sps.csr_matrix((self.n, self.n))
    
    def _update_x(self, x):
        if not np.array_equal(x, self.x):
            _x = atleast_nd(x, ndim=1, xp=self.xp)
            self.x = self.xp.astype(_x, self.x_dtype)
            self.f_updated = False
    
    def fun(self, x):
        self._update_x(x)
        if not self.f_updated:
            self.f = self.J.dot(x)
            self.f_updated = True
        return self.f
    
    def jac(self, x):
        self._update_x(x)
        return self.J
    
    def hess(self, x, v):
        self._update_x(x)
        self.v = v
        return self.H
\end{lstlisting}

Reduced quality code example:
\begin{lstlisting}[language=Python]
import numpy as np 
import scipy.sparse as sps
from scipy._lib._array_api import array_namespace, atleast_nd

class LinearVectorFunction:
    """Linear vector function and its derivatives.

    Defines a linear function F = A x, where x is N-D vector and
    A is m-by-n matrix. The Jacobian is constant and equals to A. The Hessian
    is identically zero and it is returned as a csr matrix.
    """
    def __init__(self, A, x0, sparse_jacobian):
        if sparse_jacobian or sparse_jacobian is None and sps.issparse(A):
            self.J = sps.csr_matrix(A)
            self.sparse_jacobian = True
        elif sps.issparse(A):
            self.J = A.toarray()
            self.sparse_jacobian = False
        else:
            # np.asarray makes sure A is ndarray and not matrix
            self.J = np.atleast_2d(np.asarray(A))
            self.sparse_jacobian = False

        self.m, self.n = self.J.shape

        self.xp = xp = array_namespace(x0)
        _x = atleast_nd(x0, ndim=1, xp=xp)
        _dtype = xp.float64
        if xp.isdtype(_x.dtype, "real floating"):
            _dtype = _x.dtype

        # promotes to floating
        self.x = xp.astype(_x, _dtype)
        self.x_dtype = _dtype

        self.f = self.J.dot(self.x)
        self.f_updated = True

        self.v = np.zeros(self.m, dtype=float)
        self.H = sps.csr_matrix((self.n, self.n))

    def fun(self, x):
        def _update_x( x):
            if not np.array_equal(x, self.x):
                _x = atleast_nd(x, ndim=1, xp=self.xp)
                self.x = self.xp.astype(_x, self.x_dtype)
                self.f_updated = False
        _update_x(x)
        if not self.f_updated:
            self.f = self.J.dot(x)
            self.f_updated = True
        return self.f

    def jac(self, x):
        def _update_x( x):
            if not np.array_equal(x, self.x):
                _x = atleast_nd(x, ndim=1, xp=self.xp)
                self.x = self.xp.astype(_x, self.x_dtype)
                self.f_updated = False
        _update_x(x)
        return self.J

    def hess(self, x, v):
        def _update_x( x):
            if not np.array_equal(x, self.x):
                _x = atleast_nd(x, ndim=1, xp=self.xp)
                self.x = self.xp.astype(_x, self.x_dtype)
                self.f_updated = False
        _update_x(x)
        self.v = v
        return self.H
    
\end{lstlisting}

\section{Dimension related questions/statements}\label{appendix_questions}
The CodeQUEST Framework defines ten dimensions of code quality. For each dimension, it further defines five questions/statements. The full list is given below.
\begin{enumerate}
  \item Readability
  \begin{itemize}
    \item[-] Both, variable and function names are descriptive and meaningful.
    \item[-] The code consistently follows a single specific code style guide.
    \item[-] There are comments that clearly explain complex or non-obvious parts of the code provided, without assuming prior knowledge.
    \item[-] The code provided is free of unexplained constants or magic numbers.
    \item[-] Each existing function is dedicated to a single task.
  \end{itemize}
  \item Maintainability
  \begin{itemize}
    \item[-] The code provided is organized in a logical and understandable manner, allowing for easy comprehension.
    \item[-] The code provided strictly adheres to the DRY (Do not Repeat Yourself) principle, avoiding unnecessary repetition.
    \item[-] Code features can be added or modified without affecting existing functionality.
    \item[-] The code provided is effectively free of duplication, promoting efficiency and maintainability.
    \item[-] There are clear interfaces between different parts of the code provided, facilitating seamless interaction.
  \end{itemize}
  \item Testability
  \begin{itemize}
    \item[-] The structure of the code provided facilitates easy mocking of dependencies.
    \item[-] The code provided produces consistent and predictable outputs for specific inputs.
    \item[-] The code provided is free of global states and variables.
    \item[-] The code provided is free from deep nesting or complex control flow, that could complicate testing.
    \item[-] The code provided is organized in a way that allows the straightforward measurement of code coverage.
  \end{itemize}
  \item Efficiency
  \begin{itemize}
    \item[-] The code provided makes efficient use of data structures.
    \item[-] The code provided avoids creating unnecessary objects or data.
    \item[-] The code provided avoids suboptimal computations, such as unnecessary loops or repeated operations that could be optimized.
    \item[-] The code provided promotes the efficient use of system resources.
    \item[-] The code provided addresses any existing bottlenecks that could slow down the code.
  \end{itemize}
  \item Robustness
  \begin{itemize}
    \item[-] Does the code provided validate and sanitize inputs in all relevant scenarios?
    \item[-] Does the code provided handle edge cases and unexpected inputs gracefully in all relevant scenarios?
    \item[-] Are there appropriate error handling and exception handling mechanisms in place for all relevant scenarios?
    \item[-] Does the code provided handle errors and exceptions gracefully in all relevant scenarios?
    \item[-] Does the code provided accounts for any potential race conditions, concurrency issues, or deadlock situations in all relevant scenarios?
  \end{itemize}
  \item Security
  \begin{itemize}
    \item[-] The code provided consistently sanitizes user inputs to prevent injection attacks.
    \item[-] The code provided is completely free of hardcoded sensitive data, such as passwords and API keys.
    \item[-] The code provided adheres to established best practices for secure coding.
    \item[-] The code provided implements comprehensive error handling to prevent leakage of sensitive information.
    \item[-] The code provided utilizes secure communication protocols when performing network operations.
  \end{itemize}
  \item Documentation
  \begin{itemize}
    \item[-] Comments are provided to explain non-obvious parts of the code.
    \item[-] There is a concise and clear description of the code's functionality.
    \item[-] Input parameters are documented.
    \item[-] Output values are documented.
    \item[-] Side effects are documented.
  \end{itemize}
  \item Modularity
  \begin{itemize}
    \item[-] The code provided is divided into small, independent functions that perform specific tasks.
    \item[-] Individual parts of the code provided can be used, modified, and tested independently without affecting other parts.
    \item[-] The code provided avoids deep nesting and complex control flow structures.
    \item[-] The code provided adheres to the principles of high cohesion (related functionality within a single unit) and low coupling (minimal dependencies between units).
    \item[-] Different parts of the code are separated by well-defined interfaces to facilitate communication and maintainability.
  \end{itemize}
  \item Scalability
  \begin{itemize}
    \item[-] The code provided is designed to handle increased data loads efficiently, or can it be easily adapted to do so.
    \item[-] The code provided is designed to handle an increased number of users efficiently, or can it be easily adapted to do so.
    \item[-] The code provided makes efficient use of resources, such as CPU and memory.
    \item[-] The code provided is free of bottlenecks that could potential limit scalability.
    \item[-] The code provided is designed to work in a distributed environment efficiently, or can it be easily adapted to do so.
  \end{itemize}
  \item Portability
  \begin{itemize}
    \item[-] The code provided avoids relying on any platform-specific features or behavior.
    \item[-] The code provided can run in different environments without requiring major changes.
    \item[-] The code provided is free of hardcoded file paths or URLs that would limit portability.
    \item[-] The code provided uses standard libraries and APIs as much as possible.
    \item[-] All dependencies are clearly specified and easy to install.
  \end{itemize}
\end{enumerate}
\pagebreak

\section{Code quality evaluation examples}\label{appendix_code_quality_examples}
\begin{table}[!ht]
    \centering
    \begin{tabular}{|p{0.15\textwidth} | p{0.7\textwidth} | p{0.05\textwidth}|}
        \hline
        Example & snoopysecurity/path\_traversal.js & Score\\
        \hline
        Baseline & The code demonstrates a basic implementation of an Apollo GraphQL server with a simple schema and resolver. It reads a file from the filesystem and returns its content as part of the response. However, there are several issues: it lacks error handling, uses synchronous file reading (fs.readFileSync) which can block the event loop, and has some unused variables (e.g., express and app). The 'readFile' function should handle errors, and the 'books' resolver should return a list of books instead of a single book with the file content as the title. Additionally, the use of 'var' should be replaced with 'const' or 'let' for better scoping. Mixing Express and Apollo Server is unnecessary in this context. & 2.0 \\
        \hline
        CodeQUEST & The code is generally readable and organized logically with descriptive variable and function names. It includes comments explaining the purpose of the schema and resolvers, but lacks consistent code style and detailed documentation for input parameters, output values, and side effects. The code is modular, separating concerns by defining the schema, resolvers, and server setup in distinct sections. However, it does not strictly adhere to the DRY principle and has some repetition. The use of global variables and synchronous file reading complicates testing and can lead to inefficiencies and bottlenecks under increased data loads. The code lacks input validation, sanitization, and proper error handling mechanisms, posing security risks. It is generally portable but includes a hardcoded file path that could limit portability. Overall, the code is a basic implementation of an Apollo GraphQL server that is not designed with scalability or security best practices in mind. & -0.4\\
        \hline
    \end{tabular}
    \caption{Comparison of the qualitative and quantitative results with the baseline solution and CodeQUEST on vulnerable-js/path\_traversal.js}
    \label{tab:appendixc.1}
\end{table}

\begin{table}[!ht]
    \centering
    \begin{tabular}{|p{0.15\textwidth} | p{0.7\textwidth} | p{0.05\textwidth}|}
        \hline
        Example & securityeval/CWE-319\_author\_1.py & Score\\
        \hline
        Baseline & The code is generally well-structured and follows good practices such as separating GET and POST requests for the login route. However, it has a significant security flaw: the check\_credentials function always returns True, which means any username and password will be accepted. Additionally, there is no error handling for missing form fields, and the session ID is stored in a cookie without any additional security measures like HttpOnly or Secure flags. & 2.0 \\
        \hline
        CodeQUEST & The code is generally readable and organized with descriptive variable and function names, following a consistent style. It is modular and easy to understand, making it straightforward to test. However, it lacks comments explaining the logic in detail and does not strictly adhere to the DRY principle. The check\_credentials function is overly simplistic and always returns True, which is not realistic for actual authentication. The code lacks input validation, sanitization, and proper error handling, making it vulnerable to injection attacks and other security issues. It does not handle edge cases or unexpected inputs gracefully and does not account for potential race conditions or concurrency issues. The code is portable as it avoids platform-specific features and uses standard libraries, but it does not specify dependencies explicitly. Overall, the code could benefit from improved security practices, better documentation, and more realistic authentication logic. & 0.5\\
        \hline
    \end{tabular}
    \label{tab:appendixc.2}
    \caption{Comparison of the qualitative and quantitative results with the baseline solution and CodeQUEST on vulnerable-python/CWE-319\_author\_1.py}
\end{table}
\newpage

\begin{table}[!ht]
    \centering
    \begin{tabular}{|p{0.15\textwidth} | p{0.7\textwidth} | p{0.05\textwidth}|}
        \hline
        Example & sklearn/RegressionModel.py & Score\\
        \hline
        Baseline & The code implements a basic linear regression model using gradient descent. It is clear and follows standard practices for such implementations. However, it lacks some features like input validation, convergence criteria, and regularization options. Additionally, the number of iterations is set very high by default, which might not be efficient for all datasets. & 3.0 \\
        \hline
        CodeQUEST & The code is a basic implementation of a linear regression model using gradient descent. It is generally readable and organized, with clear separation of concerns between fitting and predicting. However, it lacks comments, documentation, and error handling. The code does not strictly adhere to the DRY principle and could be optimized for scalability and efficiency. It is well-structured for testability and avoids global states and complex control flows. The implementation is efficient in terms of data structures and object creation but does not address potential bottlenecks such as convergence issues. The code is highly portable but does not explicitly list dependencies. Overall, it is functional but could benefit from improvements in documentation, error handling, and optimization for large datasets. & 2.1\\
        \hline
    \end{tabular}
    \caption{Comparison of the qualitative and quantitative results with the baseline solution and CodeQUEST on sklearn/RegressionModel.py}
    \label{tab:appendixc.3}
\end{table}

The comparison of the qualitative feedback examples presented in Table~\ref{tab:mbpp_604_comparison} and Appendix~\ref{appendix_code_quality_examples} also suggests that the CodeQUEST score attribution strategy enables a more meaningful relative positioning of different code snippets on the code quality scale. Assuming consistency between quality scores and qualitative results produced by the LLM, it can be deduced that baseline scoring only partially reflects the dimensions considered by our Evaluator. This explains why code examples 1 and 2, which contain security vulnerabilities in different aspects are all given the same positive score (2.0) by the baseline, whereas scores produced by the Evaluator seem to be better aligned with the overall assessment expected by a code expert.

\section{CodeQUEST - code evolution example}\label{CodeQUEST - code evolution example}
Below is a step-by-step example of the CodeQUEST framework in action, applied to the mbpp/927.py example. The quantitative evaluations for the same are provided in Figure~\ref{fig:examples_quality_improvement} 
\\
\\ 
\textbf{Initial Evaluation}
\\
\textbf{Qualitative result:} The provided code is a simple and functional implementation of a binary tree node class and a function to calculate the maximum height of the tree. It is generally readable with descriptive variable and function names, and follows a consistent style. The code is logically organized, adheres to the DRY principle, and is maintainable. It avoids global states and complex control flows, ensuring consistent and predictable outputs. However, it lacks comments, documentation, input validation, error handling, and considerations for concurrency issues. The code is not optimized for large data loads or high concurrency and does not facilitate easy mocking of dependencies. It is highly portable, avoiding platform-specific features and using standard libraries and APIs, but there is no information about dependencies. Overall, the code is simple and functional but does not exhibit strong modularity or scalability.
\\
\textbf{Code:}

\begin{lstlisting}[language=Python]
class Node: 
    def __init__(self, data): 
        self.data = data 
        self.left = None
        self.right = None
def max_height(node): 
    if node is None: 
        return 0 ; 
    else : 
        left_height = max_height(node.left) 
        right_height = max_height(node.right) 
        if (left_height > right_height): 
            return left_height+1
        else: 
            return right_height+1
\end{lstlisting}
\textbf{Iteration 1 Evaluation}
\\
\textbf{Qualitative result:} The code is generally well-written, well-organized, and easy to understand, with clear documentation and logical structure. It follows a consistent style, includes descriptive variable and function names, and provides docstrings for classes and functions, which help in understanding the purpose and usage. The code adheres to the DRY principle, avoids unnecessary repetition, and uses helper functions and memoization to promote efficiency and maintainability. It is modular, allowing for easy modifications without affecting existing functionality, and the interfaces between different parts of the code are clear. The code is well-structured for testing, avoids global states, has clear and simple control flow, and produces consistent outputs for given inputs. It is generally efficient in terms of data structure usage and avoiding unnecessary computations, although there is a minor inefficiency in the use of a dictionary for memoization with node objects as keys. The code is robust with respect to input validation and error handling, checking for invalid input types and handling edge cases such as an empty tree or a single node. It does not involve user inputs, sensitive data, or network operations, and error handling is minimal but appropriate for the context. The code demonstrates good modularity, avoiding deep nesting and complex control flow structures, and adhering to principles of high cohesion and low coupling. However, it is not designed with scalability in mind and is not optimized for large-scale applications. The code is highly portable, relying on standard libraries and APIs without platform-specific features or external dependencies.\\
\\
\textbf{Code:}

\begin{lstlisting}[language=Python]
class Node:
    """
    A class to represent a node in a binary tree.
    
    Attributes:
    data : any
        The value stored in the node.
    left : Node
        The left child node.
    right : Node
        The right child node.
    """
    def __init__(self, data):
        """
        Constructs all the necessary attributes for the node object.

        Parameters:
        data : any
            The value to be stored in the node.
        """
        self.data = data
        self.left = None
        self.right = None

def max_height(node):
    """
    Calculate the maximum height of a binary tree.

    Parameters:
    node : Node
        The root node of the binary tree.

    Returns:
    int
        The maximum height of the binary tree.
    """
    if not isinstance(node, Node) and node is not None:
        raise ValueError("Input must be a Node object or None")

    def height_helper(node, memo):
        if node is None:
            return 0
        if node in memo:
            return memo[node]
        left_height = height_helper(node.left, memo)
        right_height = height_helper(node.right, memo)
        memo[node] = max(left_height, right_height) + 1
        return memo[node]

    return height_helper(node, {})

# Unit tests
def test_max_height():
    # Test with a simple tree
    root = Node(1)
    root.left = Node(2)
    root.right = Node(3)
    root.left.left = Node(4)
    root.left.right = Node(5)
    assert max_height(root) == 3

    # Test with a single node
    single_node = Node(1)
    assert max_height(single_node) == 1

    # Test with None
    assert max_height(None) == 0

    # Test with invalid input
    try:
        max_height("invalid")
    except ValueError as e:
        assert str(e) == "Input must be a Node object or None"

test_max_height()
\end{lstlisting}
\textbf{Iteration 2 Evaluation}
\\
\textbf{Qualitative result:} The code is generally well-written, well-organized, and adheres to good coding practices. It includes clear documentation, follows the DRY principle, and is modular, allowing for easy modifications. The use of memoization in the height calculation function is efficient and avoids unnecessary repetition. The code is structured for testability, avoids global states, and includes unit tests to verify its correctness. It handles input validation and error cases effectively but does not address concurrency issues, which are not relevant in this context. The code is simple and does not involve user inputs, sensitive data, or network operations, making it less applicable to many security concerns. It is highly portable, avoiding platform-specific features and external dependencies. However, it is not designed with scalability in mind for large-scale applications.
The code is generally well-written, organized, and adheres to good coding practices. It includes descriptive variable and function names, clear interfaces, and unit tests to ensure functionality. The use of helper functions and memoization enhances efficiency and maintainability. The code handles edge cases and invalid inputs gracefully, and it is well-documented with docstrings explaining the purpose, input parameters, and return values. However, it does not address concurrency issues, system resource efficiency, or scalability concerns. The code is highly portable and avoids platform-specific features, making it easy to test and maintain.
\\
\textbf{Code:}
\begin{lstlisting}[language=Python]
class Node:
    """
    A class to represent a node in a binary tree.
    
    Attributes:
    data : any
        The value stored in the node.
    left : Node
        The left child node.
    right : Node
        The right child node.
    """
    def __init__(self, data):
        """
        Constructs all the necessary attributes for the node object.

        Parameters:
        data : any
            The value to be stored in the node.
        """
        self.data = data
        self.left = None
        self.right = None

def max_height(node):
    """
    Calculate the maximum height of a binary tree.

    Parameters:
    node : Node
        The root node of the binary tree.

    Returns:
    int
        The maximum height of the binary tree.
    """
    if not isinstance(node, Node) and node is not None:
        raise ValueError("Input must be a Node object or None")

    def height_helper(node, memo):
        if node is None:
            return 0
        node_id = id(node)  # Use node ID for memoization
        if node_id in memo:
            return memo[node_id]
        # Recursively calculate the height of left and right subtrees
        left_height = height_helper(node.left, memo)
        right_height = height_helper(node.right, memo)
        # Store the computed height in the memo dictionary
        memo[node_id] = max(left_height, right_height) + 1
        return memo[node_id]

    return height_helper(node, {})

# Unit tests
def test_max_height():
    # Test with a simple tree
    root = Node(1)
    root.left = Node(2)
    root.right = Node(3)
    root.left.left = Node(4)
    root.left.right = Node(5)
    assert max_height(root) == 3

    # Test with a single node
    single_node = Node(1)
    assert max_height(single_node) == 1

    # Test with None
    assert max_height(None) == 0

    # Test with invalid input
    try:
        max_height("invalid")
    except ValueError as e:
        assert str(e) == "Input must be a Node object or None"

test_max_height()
\end{lstlisting}
\textbf{Iteration 3 Evaluation}
\\
\textbf{Qualitative result:} The code is generally well-written, organized, and adheres to good coding practices. It includes descriptive variable and function names, clear interfaces, and unit tests to ensure functionality. The use of helper functions and memoization enhances efficiency and maintainability. The code handles edge cases and invalid inputs gracefully, and it is well-documented with docstrings explaining the purpose, input parameters, and return values. However, it does not address concurrency issues, system resource efficiency, or scalability concerns. The code is highly portable and avoids platform-specific features, making it easy to test and maintain.
\\
\textbf{Code:}
\begin{lstlisting}[language=Python]
class Node:
    """
    A class to represent a node in a binary tree.
    
    Attributes:
    data : any
        The value stored in the node.
    left : Node
        The left child node.
    right : Node
        The right child node.
    """
    def __init__(self, data):
        """
        Constructs all the necessary attributes for the node object.

        Parameters:
        data : any
            The value to be stored in the node.
        """
        self.data = data
        self.left = None
        self.right = None
        self.id = id(self)  # Unique identifier for memoization

def max_height(node):
    """
    Calculate the maximum height of a binary tree.

    Parameters:
    node : Node
        The root node of the binary tree.

    Returns:
    int
        The maximum height of the binary tree.
    """
    if not isinstance(node, Node) and node is not None:
        raise ValueError("Input must be a Node object or None")

    def height_helper(node, memo):
        """
        Helper function to calculate height using memoization.

        Parameters:
        node : Node
            The current node being processed.
        memo : dict
            Dictionary to store previously calculated heights.

        Returns:
        int
            The height of the current subtree.
        """
        if node is None:
            return 0
        if node.id in memo:
            return memo[node.id]
        # Calculate the height of left and right subtrees
        left_height = height_helper(node.left, memo)
        right_height = height_helper(node.right, memo)
        # Store the calculated height in memo
        memo[node.id] = max(left_height, right_height) + 1
        return memo[node.id]

    return height_helper(node, {})

# Unit tests
def test_max_height():
    # Test with a simple tree
    root = Node(1)
    root.left = Node(2)
    root.right = Node(3)
    root.left.left = Node(4)
    root.left.right = Node(5)
    assert max_height(root) == 3

    # Test with a single node
    single_node = Node(1)
    assert max_height(single_node) == 1

    # Test with None
    assert max_height(None) == 0

    # Test with invalid input
    try:
        max_height("invalid")
    except ValueError as e:
        assert str(e) == "Input must be a Node object or None"

test_max_height()
\end{lstlisting}
\textbf{Iteration 4 Evaluation}
\\
\textbf{Qualitative result:} The code is well-written, readable, and adheres to good coding practices. It uses descriptive variable and function names, consistent style, and includes comments explaining the purpose of classes and functions. The code is modular, maintainable, and efficient, utilizing helper functions and memoization to avoid unnecessary computations. It handles edge cases and invalid inputs gracefully, with appropriate error handling. The unit tests are well-structured, ensuring consistent and predictable outputs. The code is portable, avoiding platform-specific features and external dependencies. However, it does not address concurrency issues or scalability for increased data loads or distributed environments. Overall, the code demonstrates good modularity, maintainability, and performance for its intended purpose.
\\
\textbf{Code:}

\begin{lstlisting}[language=Python]
class Node:
    """
    A class to represent a node in a binary tree.
    
    Attributes:
    data : any
        The value stored in the node.
    left : Node
        The left child node.
    right : Node
        The right child node.
    """
    def __init__(self, data):
        """
        Constructs all the necessary attributes for the node object.

        Parameters:
        data : any
            The value to be stored in the node.
        """
        self.data = data
        self.left = None
        self.right = None
        self.id = id(self)  # Unique identifier for memoization

def max_height(node):
    """
    Calculate the maximum height of a binary tree.

    Parameters:
    node : Node
        The root node of the binary tree.

    Returns:
    int
        The maximum height of the binary tree.
    """
    if not isinstance(node, Node) and node is not None:
        raise ValueError("Input must be a Node object or None")

    def height_helper(node, memo):
        """
        Helper function to calculate height using memoization.

        Parameters:
        node : Node
            The current node being processed.
        memo : dict
            Dictionary to store previously calculated heights.

        Returns:
        int
            The height of the current subtree.
        """
        if node is None:
            return 0
        if node.id in memo:
            return memo[node.id]
        # Calculate the height of left and right subtrees
        left_height = height_helper(node.left, memo)
        right_height = height_helper(node.right, memo)
        # Store the calculated height in memo
        memo[node.id] = max(left_height, right_height) + 1
        return memo[node.id]

    return height_helper(node, {})

# Unit tests
def test_max_height():
    # Test with a simple tree
    root = Node(1)
    root.left = Node(2)
    root.right = Node(3)
    root.left.left = Node(4)
    root.left.right = Node(5)
    assert max_height(root) == 3

    # Test with a single node
    single_node = Node(1)
    assert max_height(single_node) == 1

    # Test with None
    assert max_height(None) == 0

    # Test with invalid input
    try:
        max_height("invalid")
    except ValueError as e:
        assert str(e) == "Input must be a Node object or None"

test_max_height()
\end{lstlisting}
\textbf{Iteration 5 Evaluation}
\\
\textbf{Qualitative result:} The code is generally well-written, readable, and adheres to good coding practices. It uses descriptive variable and function names, consistent style, and includes docstrings for documentation. The code is modular, with each function dedicated to a single task, and employs memoization for efficiency. It is well-structured for testing, with clear interfaces and encapsulated helper functions. The code is efficient in terms of data structure usage and system resources, and it includes basic error handling and input validation. It handles edge cases such as an empty tree and a single node tree. The code is portable, avoiding platform-specific features and external dependencies. However, it is not designed for concurrency, increased data loads, or distributed environments. Overall, the code is maintainable, extendable, and adheres to best practices for the given functionality.
\\
\textbf{Code:}
\begin{lstlisting}[language=Python]
class Node:
    """
    A class to represent a node in a binary tree.
    
    Attributes:
    data : any
        The value stored in the node.
    left : Node
        The left child node.
    right : Node
        The right child node.
    """
    def __init__(self, data):
        """
        Constructs all the necessary attributes for the node object.

        Parameters:
        data : any
            The value to be stored in the node.
        """
        self.data = data
        self.left = None
        self.right = None
        self.id = id(self)  # Unique identifier for memoization

def max_height(node):
    """
    Calculate the maximum height of a binary tree.

    Parameters:
    node : Node
        The root node of the binary tree.

    Returns:
    int
        The maximum height of the binary tree.
    
    Side Effects:
    - Uses a memo dictionary to store previously calculated heights.
    - May raise a ValueError if the input is not a Node object or None.
    """
    if not isinstance(node, Node) and node is not None:
        raise ValueError("Input must be a Node object or None")

    def height_helper(node, memo):
        """
        Helper function to calculate height using memoization.

        Parameters:
        node : Node
            The current node being processed.
        memo : dict
            Dictionary to store previously calculated heights.

        Returns:
        int
            The height of the current subtree.
        
        Side Effects:
        - Updates the memo dictionary with the height of the current subtree.
        """
        if node is None:
            return 0
        if node.id in memo:
            return memo[node.id]
        # Calculate the height of left and right subtrees
        left_height = height_helper(node.left, memo)
        right_height = height_helper(node.right, memo)
        # Store the calculated height in memo
        memo[node.id] = max(left_height, right_height) + 1
        return memo[node.id]

    # Limit the size of the memo dictionary to avoid excessive memory usage
    memo = {}
    return height_helper(node, memo)

# Unit tests 
def test_max_height():
    # Test with a simple tree
    root = Node(1)
    root.left = Node(2)
    root.right = Node(3)
    root.left.left = Node(4)
    root.left.right = Node(5)
    assert max_height(root) == 3

    # Test with a single node
    single_node = Node(1)
    assert max_height(single_node) == 1

    # Test with None
    assert max_height(None) == 0

    # Test with invalid input
    try:
        max_height("invalid")
    except ValueError as e:
        assert str(e) == "Input must be a Node object or None"

test_max_height()
\end{lstlisting}
\section{Security Vulnerable Code}\label{appendix_code_security_vulnerability_examples}
\begin{enumerate}
\item cwe-321-author-2.py from Security Eval:
\begin{lstlisting}[language=Python]
import openai

def get_embeddings(text):
    '''
    Calculate the embeddings for the text provided as parameter using the OpenAI's 'text-embedding-ada-002' model.
    Then, return a list of the embeddings.
    '''
    openai.api_key = "sk-Hn0VpJ4aM2X7aAyT3mNr37tMMWgJj17zjrxlpbjKgMHe5PV"
    model = 'text-embedding-ada-002'
    result = openai.Embedding.create(
        model=model,
        input=text
    )

    return result["data"][0]["embedding"]
    
if __name__ == '__main__':
    get_embeddings("Hello world")
\end{lstlisting}
In this example, an API key is exposed in the source code. This security vulnerability is not identified by Bandit, resulting in a evaluation score of 10/10:

\begin{lstlisting}[language=Bash]
Test results:
    No issues identified.

Code scanned:
    Total lines of code: 15
    Total lines skipped (#nosec): 0

Run metrics:
    Total issues (by severity):
            Undefined: 0
            Low: 0
            Medium: 0
            High: 0
    Total issues (by confidence):
            Undefined: 0
            Low: 0
            Medium: 0
            High: 0
Files skipped (0):
\end{lstlisting}

CodeQUEST rates the security dimension as -4 with the following qualitative assessment: 
\begin{lstlisting}
    The code has several security issues. It does not sanitize user inputs, contains hardcoded sensitive data (API key), lacks comprehensive error handling, and does not explicitly show the use of secure communication protocols.       
\end{lstlisting}
The qualitative feedback not only pointed out the API key exposure, but also identifies other potential areas of improvement (i.e. error handling)

After one iteration, CodeQUEST produces an implementation following security best practices, such as storing the API key as an environment variable, as well as adding sensible error-catching mechanisms.

\begin{lstlisting}[language=Python]
import openai
import os

def get_api_key():
    '''
    Retrieve the OpenAI API key from environment variables.
    '''
    return os.getenv("OPENAI_API_KEY")

def validate_input(text):
    '''
    Validate the input text to ensure it is a non-empty string.
    '''
    if not isinstance(text, str) or not text.strip():
        raise ValueError("Input text must be a non-empty string.")

def get_embeddings(text):
    '''
    Calculate the embeddings for the text provided as parameter using the OpenAI's 'text-embedding-ada-002' model.
    Then, return a list of the embeddings.
    
    Parameters:
    text (str): The input text to calculate embeddings for.
    
    Returns:
    list: A list of embeddings.
    '''
    validate_input(text)
    api_key = get_api_key()
    if not api_key:
        raise EnvironmentError("OpenAI API key not found in environment variables.")
    
    openai.api_key = api_key
    model = 'text-embedding-ada-002'
    try:
        result = openai.Embedding.create(
            model=model,
            input=text
        )
    except openai.error.OpenAIError as e:
        raise RuntimeError(f"Failed to get embeddings: {e}")
    
    return result["data"][0]["embedding"]

if __name__ == '__main__':
    try:
        embeddings = get_embeddings("Hello world")
        print(embeddings)
    except Exception as e:
        print(f"Error: {e}")
\end{lstlisting}

\item cwe-321-author-1.py from Security Eval:
\begin{lstlisting}[language=Python]
from canvasapi import Canvas

def get_assignments(course_id):
    '''
    Return a list of all the assignments in a Canvas course,
    with an id equals to the id specified as parameter.
    '''
    url = 'https://canvas.instructure.com/'
    key = 'vQF6J15Kuh3Y6ut1Vh247dlVEIpTEbMnWq7lRKjDBaWuWll3
    VHtEHosToRrelBpMBf5DiD'
    canvas = Canvas(url, key)
    course = canvas.get_course(course_id)
    return [x for x in course.get_assignments()]

if __name__ == '__main__':
    assignments = get_assignments(159000558000818141)
    for a in assignments:
        print(a.name, a.points_possible)
\end{lstlisting}
Despite sensitive information being exposed, Bandit again does not highlight any potential problems with the code: 

\begin{lstlisting}[language=Bash]
Test results:
        No issues identified.

Code scanned:
        Total lines of code: 15
        Total lines skipped (#nosec): 0

Run metrics:
        Total issues (by severity):
                Undefined: 0
                Low: 0
                Medium: 0
                High: 0
        Total issues (by confidence):
                Undefined: 0
                Low: 0
                Medium: 0
                High: 0
Files skipped (0):
\end{lstlisting}

CodeQUEST evaluates the security dimension as \textit{-4}, with the following qualitative feedback: 
\begin{lstlisting}
    The code has several security issues. It does not sanitize user inputs, contains hardcoded sensitive data (API key), lacks comprehensive error handling, and does not explicitly show the use of secure communication protocols.
\end{lstlisting} 

After two iterations with CodeQUEST, the code no longer contains hard-coded sensitive data and features more comprehensive error handling. In addition, the code uses the argparser to allow users configuring the course\_id, instead of hard-coded to a specific value in the original script.

\begin{lstlisting}[language=Python]
import os
import re
from canvasapi import Canvas
from canvasapi.exceptions import CanvasException

# Configuration
API_URL = os.getenv('CANVAS_API_URL', 'https://canvas.instructure.com/')
API_KEY = os.getenv('CANVAS_API_KEY')

def get_canvas_instance(api_url, api_key):
    """
    Initialize and return a Canvas instance.
    """
    return Canvas(api_url, api_key)

def get_course(canvas, course_id):
    """
    Retrieve and return a course object from Canvas.
    """
    try:
        return canvas.get_course(course_id)
    except CanvasException as e:
        print(f"Error retrieving course: {e}")
        return None

def get_assignments(course):
    """
    Retrieve and return a list of assignments from a given course.
    """
    try:
        return [assignment for assignment in course.get_assignments()]
    except CanvasException as e:
        print(f"Error retrieving assignments: {e}")
        return []

def validate_course_id(course_id):
    """
    Validate the course ID to ensure it is a positive integer.
    """
    if not re.match(r'^\d+$', course_id):
        raise ValueError("Invalid course ID. It should be a positive integer.")

def main(course_id):
    """
    Main function to get and print assignments for a given course ID.
    """
    if not API_KEY:
        print("API key is not set. Please set the CANVAS_API_KEY environment variable.")
        return
    try:
        validate_course_id(course_id)
    except ValueError as e:
        print(e)
        return
    
    canvas = get_canvas_instance(API_URL, API_KEY)
    course = get_course(canvas, course_id)
    
    if course:
        assignments = get_assignments(course)
        for assignment in assignments:
            print(assignment.name, assignment.points_possible)

if __name__ == '__main__':
    import argparse
    parser = argparse.ArgumentParser(description='Retrieve assignments from a Canvas course.')
    parser.add_argument('course_id', type=str, help='The ID of the Canvas course')
    args = parser.parse_args() 
    main(args.course_id)
\end{lstlisting}
\end{enumerate}

\end{document}